\documentclass[11pt]{cernrep}
\usepackage{graphicx}
\usepackage{here}
\newcommand{\Tr}{{\rm \, Tr \!}}    
\newcommand{\g}{\gamma}
\newcommand{\da}{\dagger}  

\begin{document}
 \title{RECENT RESULTS ON THE TRANSVERSE LATTICE}
\author{S. Dalley~$^{a}$, B. van de Sande~$^{b}$}
\institute{$^{a}$~Department of Physics, 
University of Wales Swansea, United Kingdom\\
$^{b}$~Department of Physics, Geneva College, Pennsylvania}

\maketitle
\begin{abstract}
We review recent progress 
of field theory on a transverse lattice including: 
realistic calculations of pion structure in large-$N_c$ QCD;
fermion doubling and chiral symmetry; a new strong coupling limit; 
supersymmetry on the transverse lattice; 
lightcone zero-mode analysis.

\end{abstract}

\section{INTRODUCTION}

The transverse lattice formulation of gauge theories 
\cite{bard} has been developed in recent years as a tool
for hadronic physics \cite{rev}. It has particularly been applied
to phenomenological studies of pure glue and 
glueballs \cite{bv,mat1} and mesons \cite{mat2,sd1,mat3} 
in the large $N_c$ limit on coarse
lattices, in the context of the colour-dielectric expansion. 
More recently, as well as refinements of those studies \cite{sd2},  
a number a new directions have been explored for the
transverse lattice \cite{hari1,rata,sdlcq2,past}.

\section{THE PION AND BEYOND}

We first introduce the basic structure common to all
transverse lattice approaches.
In spacetime we introduce a square 
lattice of spacing $a$
in the `transverse' directions ${\bf x}=\{x^1,x^2\}$ 
and a continuum in the $\{x^0,x^3\}$ directions.
The lightcone coordinates are $x^{\pm} = (x^0 \pm x^3)/\sqrt{2}$ and
$x^+$ is treated as canonical time; we use indices $r,s \in \{1,2\}$
and $\alpha, \beta \in \{+,-\}$.
$SU(N)$ gauge field 
degrees of freedom are represented by continuum 
Hermitian gauge potentials 
$A_{\alpha}({\bf x},x^+,x^-)$ and 
$N$ x $N$  link matrices $M_r({\bf x},x^+,x^-)$. 
$A_{\alpha}({\bf x})$ and Dirac fermions $\Psi({\bf x})$ reside on the plane
${\bf x} = {\rm constant}$, while $M_r({\bf x})$ is associated
with a link from ${\bf x}$ to ${\bf x} + a \hat{\bf r}$, where
$\hat{\bf r}$ is a unit vector in direction $r$. $M_r({\bf x})^{\dagger}$
goes from ${\bf x} + a \hat{\bf r}$ to ${\bf x}$.
These variables map under transverse lattice gauge
transformations $V({\bf x},x^+,x^-) \in SU(N)$ as
\begin{eqnarray}
        A_{\alpha}({\bf x}) & \to & V({\bf x}) A_{\alpha}({\bf x}) 
        V^{\dagger}({\bf x}) + 
{\rm i} \left(\partial_{\alpha} V({\bf x})\right) 
        V^{\dagger}({\bf x})  \nonumber \\
        M_r({\bf x}) &  \to & V({\bf x}) M_r({\bf x})  
        V^{\dagger}({\bf x} + a\hat{\bf r})\nonumber \\ 
        \Psi({\bf x}) & \to & V({\bf x})\Psi({\bf x})  \ .
\end{eqnarray}
From these fields one can write down gauge invariant actions of the form
\begin{eqnarray}
L & = &  \sum_{{\bf x}} \int dx^- \sum_{\alpha, \beta = +,-}
\sum_{r=1,2} 
-{1 \over 2 G^2} \Tr \left\{ F^{\alpha \beta}({\bf x}) 
F_{\alpha \beta}({\bf x}) \right\}
 \nonumber
\\
&& + \Tr\left\{\overline{D}_{\alpha}M_r({\bf x}) (\overline{D}^{\alpha}
M_r({\bf x}))^{\dagger}\right\} 
\nonumber \\
&& - \mu_{b}^2  \Tr\left\{M_r M_r^{\da}\right\}
 + {\rm i} \overline{\Psi} 
\g^{\alpha} (\partial_{\alpha} + {\rm i} A_{\alpha}) \Psi - \mu_f
\overline{\Psi}\Psi 
\nonumber\\
&& +  {\rm i} \kappa_a \left( \overline{\Psi}({\bf x}) \g^{r} M_r({\bf x})
 \Psi({\bf x} + a \hat{\bf r}) 
- \overline{\Psi}({\bf x}) \g^{r} M_{r}^{\da}({\bf x}- a \hat{\bf r}) 
\Psi({\bf x} - a \hat{\bf r})
\right)\nonumber\\
&&
+ \kappa_s \left( \overline{\Psi}({\bf x}) M_r({\bf x})
 \Psi({\bf x} + a \hat{\bf r})+\overline{\Psi}({\bf x}) M_{r}^{\da}({\bf
x}- 
a \hat{\bf r})
 \Psi({\bf x} - a \hat{\bf r})\right) - \nonumber \\ 
&& - {\beta\over N_c a^2} \sum_{r \neq s} 
  \Tr\left\{ M_{r} ({\bf x}) 
M_{s} ({\bf x} + a \hat{\bf r})
M_{r}^{\da}
({\bf x} + a \hat{\bf s} )  
M_{s}^{\da}({\bf x}) \right\} + \cdots \ ,
\label{ferlag}
\end{eqnarray}
where $F^{\alpha \beta}({\bf x})$ is the continuum field strength in the
$(x^0,x^3)$ planes at each ${\bf x}$,
\begin{equation}
\overline{D}_{\alpha}M_r({\bf x}) = (\partial_{\alpha} + 
{\rm i} A_{\alpha} ({\bf x}))
        M_r({\bf x})-  {\rm i} M_r({\bf x})   A_{\alpha}({{\bf x}+a
 \hat{\bf r}})
\end{equation}
and $\cdots$ indicates other gauge-invariant terms, typically  at higher order
in powers of the fields $M_r$ and $\Psi$.

If we allow only kinetic terms that are quadratic then, in the lightcone
gauge $A_{-} = 0$, 
it is straightforward to formally 
derive the most general light-cone  
hamiltonian allowed for this system. One way to truncate this most
general hamiltonian for practical applications
is to work on coarse lattices, using the
colour-dielectric expansion. This assumes the link variables $M_r$
are highly disordered and take values in the 
space of all complex matrices. This has the advantage that linearized
variables are easy to quantize and hadrons are small in lattice units,
but leads to complicated effective hamiltonians.
Alternatively,
as the continuum limit $a \to 0$ is approached, the link variables
are forced into the $SU(N_c)$ group manifold and all but a few
terms in the action are irrelevant. However, it is difficult to
quantize these non-linear variables and hadrons are large in 
lattice units.
\begin{figure}
\hspace{-50mm}  
\centering
\includegraphics[width=19cm]{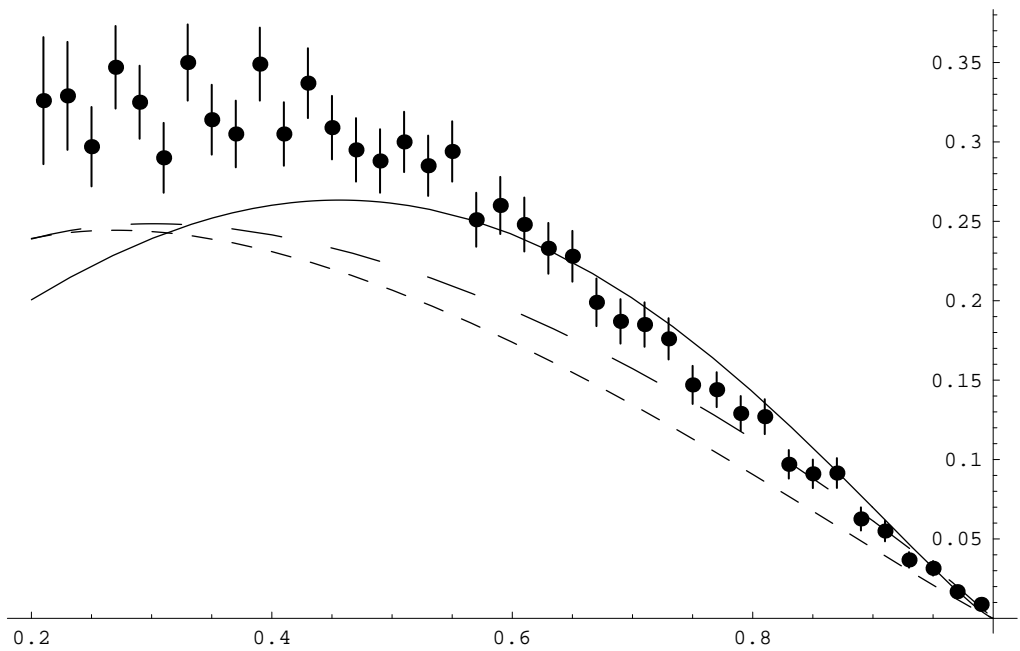} 
\vspace{-50mm} 
\caption{Valence distribution functions  $x V_{\pi}(x)$ compared
to pion-nucleon Drell-Yan data. 
Solid line: transverse
lattice result evolved to $6.6$ GeV. Data points: E615 experiment \cite{e615}.
Short-dashed line: NA10 experiment fit to $x^{\alpha} (1-x)^{\beta}$ form 
\cite{na10}.
Long-dashed line: NA3 experiment  fit to $x^{\alpha} (1-x)^{\beta}$ form 
\cite{na3}.
\label{fig1}}
\end{figure}

Recent work using the colour-dielectric expansion
combined with DLCQ
has allowed realistic calculations of the
structure of the pion at large $N_c$ \cite{sd2,sd3}.
The eigenstates of the lightcone hamiltonian are in the form of
wavefunctions that can be directly related to many experimentally
accessible observables (modulo some possible problems with final state
interactions \cite{stan}). The structure
function  $V_{\pi}(x)$, elastic form factor $F_{\pi}(Q^2)$
and distribution amplitude $\phi_{\pi}(x)$ of the pion are
in reasonable agreement with experimental data
for even the simplest truncation of the colour-dielectric
expansion.
There is an interesting comparison of these results with other
theoretical techniques discussed at this workshop. Chiral quark
models, which have an essentially structureless pion at low
resolution scales, can produce even closer agreement with Drell-Yan data
on $V_{\pi}(x)$ (see talk of Ruiz Arriola and Ref.~\cite{ruiz}). 
Dyson-Schwinger calculations
are somewhat in  disagreement with data, but seem more consistent with
perturbative QCD arguments for the $x \to 1$ behaviour (see talk of Roberts
and Ref.~\cite{ds}). 
It is
important  to clear up this discrepancy --- who is right, if anyone?

\begin{figure}
\centering
\includegraphics[width=19cm]{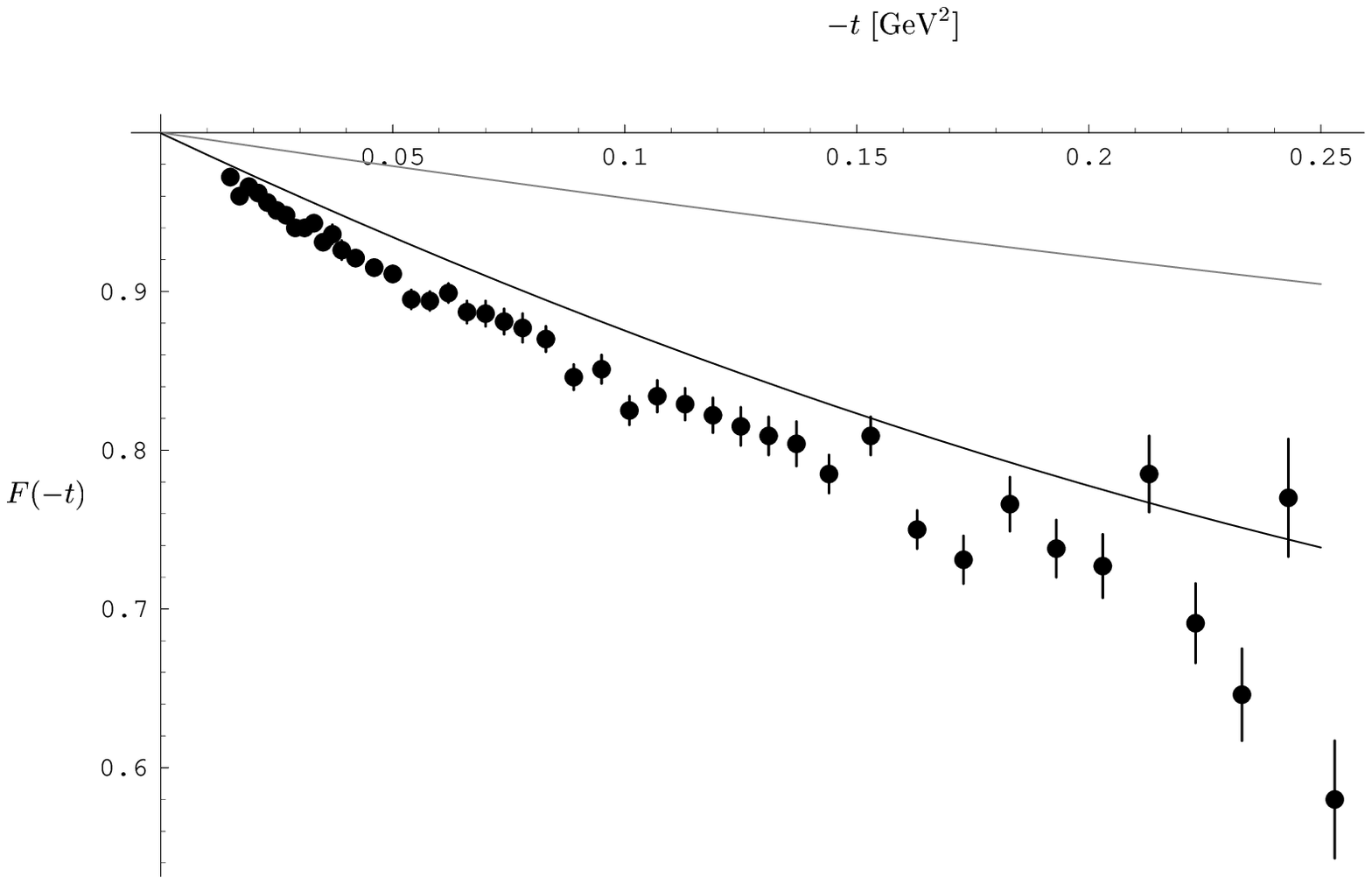}
\vspace{-100mm}
\caption{Pion form factor $F(Q^2)$. Solid curves are the transverse
lattice result, gray curve for wavefunctions calculated
within a one-link truncation, black curve for a three-link truncation. 
The experimental data
points are from Ref.~\cite{exp}.
\label{fig2}}
\end{figure}

\begin{figure}
.\vspace{-100mm}
\centering
\includegraphics[width=19cm]{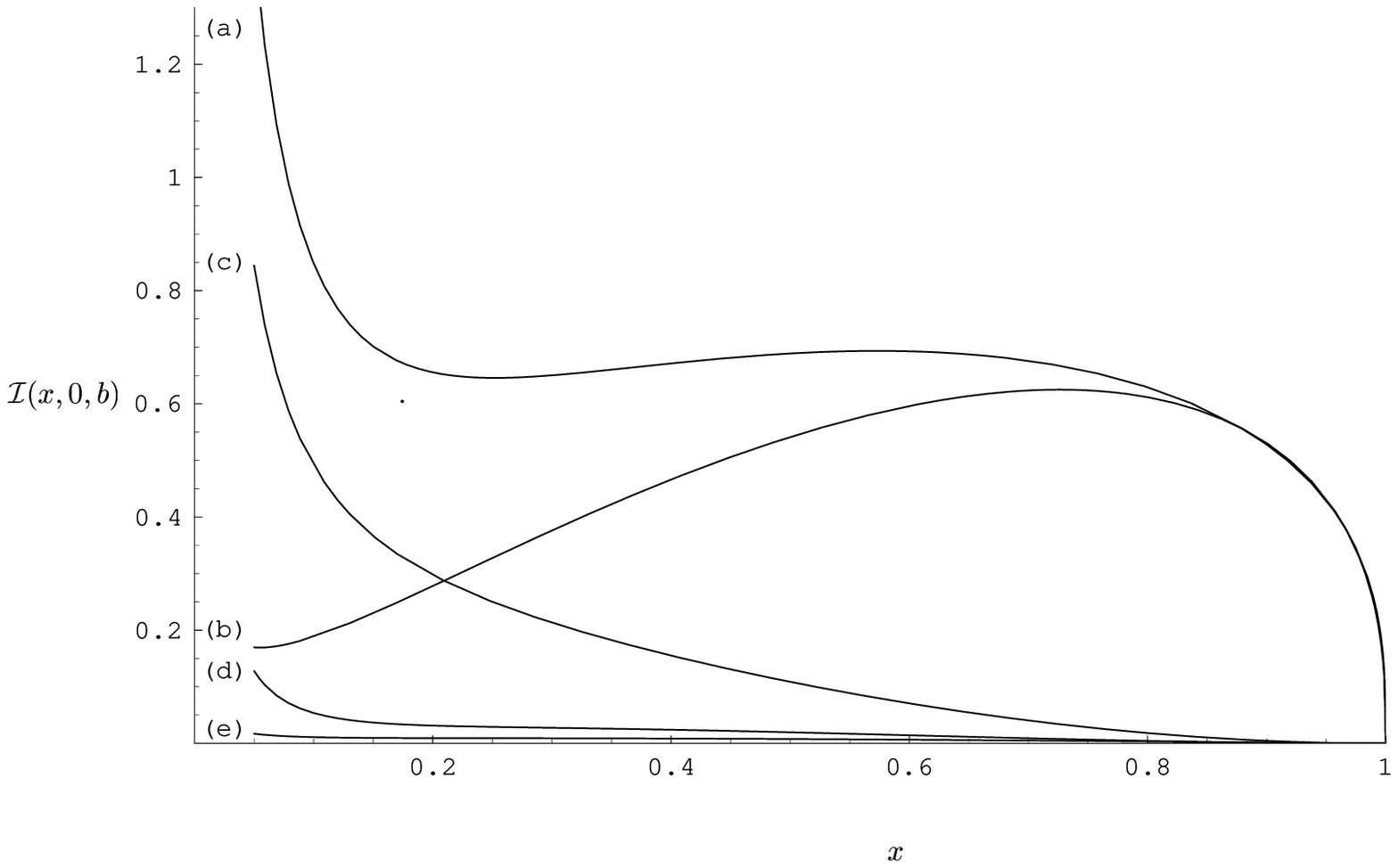}\\
\vspace{-70mm}
\caption{Impact parameter dependent valence quark distributions of the
pion. (a) The full distribution $V(x)$, which  sums 
over all impact parameters ${\bf b}$. (b)-(e) The sum of contributions 
at impact parameters ${\bf b}$ along  lattice axes such that 
$b= | {\bf b}| =0, a, 2a, 3a$ respectively. 
\label{fig3}}
\end{figure}

The corresponding
distribution of valence quark helicity, {\em i.e.}\ projection of quark spin on
the direction of motion of the fast-moving hadron, was calculated for 
the first time in the pion \cite{sd2}.
It was found to have the same shape as $V(x)$ but there is a large 
probability to find the quark and anti-quark helicities aligned
with one another, which is completely different from the picture
of the non-relativistic quark model. 
To investigate simultaneously the transverse and longitudinal structure
one can use the impact parameter dependent parton distributions
${\cal I}(x,0,{\bf b})$, where ${\bf b}$ is impact parameter, that measure
the probability of finding a quark with given Bjorken $x$ and transverse
position ${\bf b}$. 
These were calculated for the transverse lattice \cite{sd3} at  
particular lattice values of ${\bf b}$, and are shown in  Figure~\ref{fig3}.
This corresponds to a
sharp fall-off of the hadron wavefunction in transverse space at a 
particular ($x$-dependent) radius.
The talk of Broniowski in these proceedings and Ref.~\cite{bra}
shows that chiral quark
models predict something  qualitatively similar. It should be possible to 
calculate this in the Dyson-Schwinger approach also.

It is a challenge to extend this work to the baryons and other
mesons.
It the case of baryons, since  large $N_c$ no longer restricts
the number of quarks, one needs a suitable truncation of the Fock
space commensurate with the colour-dielectric expansion. In the case
of mesons, one needs to understand particularly
how the pion-rho splitting arises. The coarse lattice hamiltonian
contains spin flip operators, involving link-field emission,
that do split the pion and rho, and are
presumably a consequence of spontaneous chiral symmetry breaking,
but at the
expense of also splitting the rho multiplet itself {\em i.e.}\ 
breaking Lorentz invariance. 
One needs further operators that improve both explicit chiral
and Lorentz symmetry. One possibility allowed is four-fermi
interactions, since these are renormalisable with repsect 
to the two continuum dimensions $x^{\alpha}$. They have the potential
to split the pion from the rho without splitting the rho iteslf \cite{mat4}.
These  would bring the transverse lattice wavefunctions
to a form more closely resembling those of the chiral quark
models; in particular, they would  not vanish at Bjorken $x = 0 , 1$.

\section{FERMION DOUBLING AND CHIRALITY}

As with other lattice formulations, the transverse lattice
suffers from the fermion doubling problem. This has been
recently studied in detail by Chakrabarti {\em et al.} \cite{hari1}. 
They find that
with symmetric lattice derivatives, in the free field limit fermions
on even and odd
transverse lattice sites decouple.
As a result, four species of fermions appear on the (two-dimensional)
transverse lattice as excitations around zero transverse momentum.
This is quite different from Euclidean lattice theory, where doublers
have at least one momentum component near the edge of the Brillouin
zone. They propose a staggered fermion formulation on the lightfront
transverse lattice that eliminates two of the species, with the
other two interpreted as different flavours. 
Previous coarse transverse lattice
calculations have used Wilson fermions, which also removes doublers 
\cite{mat2}.

Another possibility, discussed in ref.\cite{hari1}, is to use separate
forward and backward derivatives. Introducing the left and right
moving components of the fermion field $\psi^{\pm} = \gamma^{\mp} 
\gamma^{\pm} \Psi$, the transverse gauge kinetic term is
\begin{equation}
i \overline{\psi}^{-} \gamma_{r} D_{r}^{\rm f} \psi^+
+ i \overline{\psi}^{+} \gamma_{r} D_{r}^{\rm b} \psi^- \ ,
\end{equation}
where
\begin{eqnarray}
D_{r}^{\rm f} \psi ({\bf x}) \equiv {1 \over a} [M_{r}({\bf x}) \psi({\bf x}
+ a \hat{\bf r}) - \psi({\bf x})] \ , \nonumber \\
D_{r}^{\rm b} \psi ({\bf x}) \equiv {1 \over a} [\psi({\bf x})
-M_{r}^{\dagger}({\bf x}- a \hat{\bf r}) \psi({\bf x}- a \hat{\bf r})] \ .
\end{eqnarray}
In the free field limit, for bare fermion mass $m$ with transverse
momentum ${\bf k } = (k_1, k_2)$, 
this leads to an invariant mass
\begin{equation}
{\cal M}^2 = m^2 + { 4 \over a^2} \sum_{r} \sin^2 {k_r a \over 2}
\pm {4m \over a} \sqrt{\sum_{r} \sin^4 {k_r a \over 2}} \  .
\end{equation}
${\cal M}^2 = m^2$ only for the case ${\bf k} = 0$ and there
is no doubling. (The last term above occurs due to a helicity
flip term in the hamiltonian). 

In all the above cases the question of explicit chiral symmetry
breaking arises; the usual theorems lead one to expect it
if doublers are removed. One must distinguish between conventional
chiral transformations, which act on $\Psi$, and lightcone
chiral transformations that act on the physical component
$\psi^+$ only. It is often the case that the latter can be
a symmetry even in theories that are obviously not chirally
symmetric in the usual sense (e.g. free massive fermions), and may
even be  useful for classifying hadrons in the quark model \cite{must}.
How to check for conventional explicit chiral symmetry? Chakrabarti
{\em et al.} suggest an `even-odd' helicity flip transformation on the
physical component
\begin{equation}
\psi^+(x^1, x^2)  \to (\gamma_1)^{x_1} (\gamma_2)^{x_2} \psi^+ (x_1, x_2)
\end{equation}
that is always violated in the cases above. Alternatively one might
try to directly check conservation of the usual chiral current \cite{sd2}. 
Conservation of the `1+1' current
\begin{equation}
j_{5}^{\alpha} = \sum_{\bf x} \overline{\Psi} \gamma_5 \gamma^{\alpha} \Psi
\end{equation}
is a necessary condition for conservation of the four-dimensional
current. One cannot check  $\partial_{\alpha} 
j_{5}^{\alpha} = 0$ directly because it contains a bad current
$j^-$ that leads to normal-ordering ambiguities. 
Suppose however we 
consider the vacuum-to-one-pseudoscalar-meson
matrix element\footnote{This approach 
arose in discussions with M. Burkardt during the
workshop}:
\begin{equation}
\langle 0|j_{5}^{\alpha}(0)|{\rm meson}(P^+, {\bf P})\rangle
= f P^{\alpha} \delta({\bf P}) \label{cc}
\end{equation}
The form of the RHS follows from exact 1+1 symmetries. 
$\partial_{\alpha} j_{5}^{\alpha} = 0$  implies that
either $P^- = 0$ or $f = 0$, depending on the meson. The 
first condition is the massless
pion condition. In addition there are an
infinite number of further conditions --- excitations of the pion
must have zero decay constant $f=0$. 
By tuning irrelevant operators, for example within the
colour-dielectric expansion, one might hope to gradually
realise these conditions. Other operators that improve
Lorentz symmetry might then gradually move  the conditions 
towards sufficiency for chiral symmetry. This is an interesting idea
yet to be attempted in practice.

Some applications of the asymmetric derivatives to meson calculations
appeared after the workshop \cite{hari2}.

\section{STRONG COUPLING LIMIT}
An interesting limit of transverse lattice gauge theory has been
formulated by Patel and Ratabole \cite{rata}, 
which is an analogue of the strong coupling
limit in conventional lattice theories. They take the link-field
$M_r$  in the $SU(N)$ group
manifold, as in the continuum limit.  $M_r$ can be totally
disordered if one turns off all its gauge and self interactions and its
kinetic term. Provided
the coupling to fermions is linear in $M_r$, which is the case with
the simplest lattice discretization of the Wilson-Dirac operator, 
the link
field can be integrated out of the path integral exactly at
large $N_c$. This allows derivation of the effective action, quark
propagator, chiral condensate, and
a boundstate equation for mesons
consisting of the 't Hooft equation modified by a meson
transverse hopping via four-fermi interaction:
\begin{eqnarray}
\left[ {\cal M}^2 - {m^2 - {g^2 \over \pi} \over x(1-x)} \right] \Phi(x)
& = & {1 \over 2x(1-x)} \left[ {m^2 - {g^2 \over \pi} \over 2x} \gamma^+
+x\gamma^- + m + {g^2 \gamma^+ \over 2 \pi x} \right]  \nonumber \\
&& \times \int {dy \over 2\pi} \left\{g^2 {\rm P} {\gamma^+ \Phi(x) \gamma^+
\over (x-y)^2} -{\rm i}\kappa^2 \left[2\Phi(x) + \sum_{n=1}^{2} \gamma^n
\Phi(x) \gamma^n \right] \right\} \nonumber \\
&& \times \left[-\left({{\cal M}^2 \over 2} - {m^2 - {g^2 \over \pi} 
\over 2x}\right) \gamma^+ -(1-x)\gamma^- +m - {g^2 \gamma^+ \over 2 \pi
(1-x)}\right] \ .
\end{eqnarray}
The above equation is for states of overall zero transverse momentum,
$m$ is the bare mass, $g$ the longitudinal gauge coupling, $\kappa$ the
Dirac hopping parameter on the lattice, and the Wilson parameter $r=1$.
The wavefunction $\Phi$ in this case is a matrix in Dirac space and
covers all the possible spin structures of the quark anti-quark
pair. This equation has yet to be solved in detail.

A possible generalisation of this idea would treat the link 
fields $M_r$ as complex matrices, so that they are already disordered,
 and include also a potential $V[M]$ localised on each link
that does not propagate links in either transverse or longitudinal
directions. Integrating out the link variables would produce a 
class of generalisations of the 't Hooft model with transverse hopping.

\section{SUPERSYMMETRY}

It is known that DLCQ in 2 dimensions preserves supersymmetry \cite{sdlcq1}.
A recent attempt has been made to extend this to higher dimensions
using a transverse lattice. 
Of course,  one may not  expect to recover exact supersymmetry,
but Harada and Pinsky \cite{sdlcq2}
have shown, using the SDLCQ technique, that
one may preserve half  of an $N=(1,1)$ supersymmetry in $2+1$ dimensions.
The fermion $\Psi$ is now in the adjoint representation. In the
lightcone SUSY algebra
\begin{equation}
\{ Q^{\pm} , Q^{\pm} \} = 2 \sqrt{2} P^{\pm} \ , \ 
\{ Q^{+} , Q^{-} \} = 2  {\bf P}
\end{equation}
they propose a charge $Q^-$ that preserves gauge invariance, whose square
defines the hamiltonian $P^-$:
\begin{equation}
Q^- = 2^{3/4} g  \sum_{\bf x} \int dx^- {\rm Tr} \{ J^+({\bf x}) 
{1 \over \partial_{-}} \psi^+({\bf x})\}
\end{equation}
\begin{equation}
J^+({\bf x}) = {{\rm i} \over 2 g^2 a^2}\left(M_{r}({\bf x}) 
\stackrel{\leftrightarrow}{\partial}_{-} 
M_r^{\dagger}({\bf x})  + M_r^{\dagger}({\bf x} - a \hat{\bf r})
\stackrel{\leftrightarrow}{\partial}_{-} M_r({\bf x} - a \hat{\bf r})
   \right) + 2 \psi^+ ({\bf x}) \psi^+({\bf x})
\end{equation}
(In this case $r=1$ is the single transverse direction.)
The continuum limit $a \to 0$ of $ P^- \sim (Q^-)^2$ reproduces
the correct Super Yang-Mills result. The existence of $Q^-$
 ensures fermion-boson degeneracy
in the massive spectrum, at least for ${\bf P}= {\bf 0}$,
but the absence of a suitable
gauge-invariant $Q^+$ means that the number of massless states
do not match. Numerical diagonalisations of the lightcone
hamiltonian problem
exhibit interesting features, such as a winding mode spectrum
around compact transverse directions that  varies inversely
with winding number. This seems to be a feature of the exact
supersymmetry in this case since one might naively expect mass to increase
with winding number, as the gauge strings get longer. 
It may be that the limit of small soft
SUSY breaking gives a different result from working exactly at the
SUSY point.

\section{ZERO MODES}
It is well known  in the context of the DLCQ regulator, which makes
$x^-$ periodic, that it is not possible to completely reach the lightcone
gauge $A_- = 0$. The $x^-$ zero mode of $A_-$ remains as a dynamical 
degree of freedom. In all explicit transverse lattice calculations
performed so far, this zero mode has simply been dropped as a 
dynamical approximation. Paston {\em et al.}\
have studied the inclusion of this zero mode in the formalism \cite{past}.
The zero mode $A_{-}^o$ can be diagonalised in colour space by 
a residual gauge choice, then the
 link fields are expanded in eigen modes of the operator
\begin{equation}
D_- M_{r} ({\bf x})= 
[\partial_{-} -i A_{-}^o({\bf x}) + i A_{-}^o({\bf x} - a \hat{\bf r})]
M_{r} ({\bf x})
\end{equation}
The state space can then be written as
 a product of the action of the above eigen modes
on a Fock vacuum $| 0 \rangle$
and  normalisable functionals $F[A_{-}^o]$
of classical functions $A_{-}^o({\bf x})$.
It is not yet clear whether this causes any substantive
effect on physical observables compared to a calculation based
on the Fock vacuum $| 0 \rangle$ alone with suitable renormalisation
of couplings.

\section*{ACKNOWLEDGEMENTS}

The work of SD is supported by PPARC 
grant PPA/G/0/2002/00470.
The work of BvdS is supported by NSF grant PHY-0200060.

\end{document}